\documentclass[%
superscriptaddress,
nofootinbib,
 amsmath,amssymb,
 aps,
]{revtex4-2}

\usepackage{graphicx}
\usepackage{dcolumn}
\usepackage{bm}
\usepackage{hyperref}
\usepackage{url}
\usepackage[mathlines]{lineno}

\begin{document}

\title{Thickness effect on superconducting properties of niobium films for radio-frequency cavity applications}

\author{Antonio Bianchi}\email{Corresponding author \\Email address: antonio.bianchi@mi.infn.it}
\affiliation{INFN, Milan, Italy}

\author{Marco Bonura}
\affiliation{Department of Quantum Matter Physics, University of Geneva, Switzerland}

\author{Carlota P. A. Carlos}
\affiliation{Department of Quantum Matter Physics, University of Geneva, Switzerland}
\affiliation{CERN, Geneva, Switzerland}

\author{Stewart Leith}
\affiliation{CERN, Geneva, Switzerland}

\author{Guillaume Rosaz}
\affiliation{CERN, Geneva, Switzerland}

\author{Carmine Senatore}
\affiliation{Department of Quantum Matter Physics and Department of Nuclear and Particle Physics, University of Geneva, Switzerland}

\author{Walter Venturini Delsolaro}
\affiliation{CERN, Geneva, Switzerland}

\begin{abstract}

Niobium-coated copper radio-frequency cavities are cost-effective alternatives to bulk niobium cavities, given the lower material costs of copper substrates and their operation in liquid helium at around 4.2 K. However, these cavities historically exhibited a gradual degradation in performance with the accelerating field. This phenomenon, not yet fully understood, limits the application of niobium thin film cavities in accelerators where the real-estate gradient needs to be maximized. Recent studies on niobium films deposited on copper using high power impulse magnetron sputtering (HiPIMS) technique show promising results in mitigating the performance degradation of niobium thin film radio-frequency cavities.

This paper examines the effect of film thickness on the superconducting properties of niobium films deposited on copper using HiPIMS. The study provides insights into how the critical temperature, transition width, lower and upper critical fields, and critical current density vary with the film thickness. Increasing the thickness of niobium films deposited through HiPIMS is found to enhance superconducting properties and reduce densities of defects and structural irregularities in the crystalline lattice. This shows potential for enhancing overall performance and potentially mitigating the observed performance degradation in niobium thin film radio-frequency cavities. Additionally, the Ivry's scaling relation among critical temperature, thickness, and sheet resistance at the normal state appears applicable to niobium films up to approximately 4 $\mu$m. This extends the previously confirmed validity for niobium films, which was limited to around 300 nm thickness.


\end{abstract}

\maketitle


\section{\label{sec:level1}INTRODUCTION}

Niobium films deposited on various substrates are currently being studied for several technological applications. These applications include detectors of single photons \cite{semenov2003, friedrich2008}, millimeter-wave receivers \cite{jia2019, anischenko1993}, sub-millimeter-wave mixers \cite{karasik1997, dierichs1993}, and superconducting quantum interference devices (SQUIDs) \cite{fujimaki1992, bouchiat2001}. Niobium films are even employed in creating superconducting qubits for the advancement of quantum computers, relying on their elevated robustness and chemical stability \cite{lisenfeld2007, bennett2007}. Furthermore, these niobium films find applications in constructing superconducting radio-frequency (SRF) cavities for particle accelerators.

Niobium-coated copper cavities were introduced in the early 1980s \cite{benvenuti1984niobium} as a promising alternative to bulk niobium cavities \cite{calatroni2006, venturini2023}. One of the advantages of employing niobium thin film technology for SRF cavities lies in the cost reduction achieved by utilizing copper in place of niobium. This type of cavities was adopted at CERN for the Large Electron Positron collider (LEP-II) and, then, for the Large Hadron Collider (LHC). At INFN-LNL, niobium-coated copper cavities were employed for the Acceleratore Lineare Per Ioni (ALPI) machine. They were also utilized in various other accelerator facilities and projects, such as ELETTRA, SOLEIL, and SLS \cite{calatroni2006}. More recently, this technology was once again used for SRF cavities in the High Intensity and Energy Isotope Separator On Line Device (HIE-ISOLDE) linac at CERN.

The performance of SRF cavities is determined by the surface resistance of the superconductor present in their internal surface. At an operational temperature well below the superconducting transition temperature, the surface resistance $R_{s}$ of a superconductor in the radio-frequency regime can be expressed as the sum of two terms: a temperature-dependent term, which depends on the fraction of unpaired carriers, and a second one, usually referred to as residual surface resistance, which is typically associated with imperfections, interstitial impurities and atomic size defects in the crystalline lattice \cite{padamsee}. Both terms of $R_{s}$ depend on various parameters, as extensively discussed in the literature \cite{bonin1996, padamsee}. An in-depth discussion is not the subject of this paper; here, we only emphasize that the superconducting properties play a fundamental role in determining the surface resistance of a superconductor. 

While niobium thin film cavities provide certain advantages, especially for large cavities operating at low frequencies \cite{bonin1996}, it is important to note that this type of cavities can exhibit serious performance degradation issues with increasing radio-frequency field \cite{benvenuti1999, aull2015, miyazaki2019two, vegacid2023}, preventing their use in accelerators where the real-estate gradient needs to be maximized. This is the so-called $Q$-slope problem, which occurs due to the increase in the surface resistance of the niobium films when exposed to an increasing radio-frequency field, resulting in higher power dissipation. Recent studies have reported a significant reduction of the $Q$-slope phenomenon and the residual surface resistance in niobium films deposited on copper using the high power impulse magnetron sputtering (HiPIMS) technique. These niobium films exhibit similar surface resistance values compared to those obtained on bulk niobium surfaces \cite{arzeo2022}. Using the HiPIMS deposition technique, promising results are obtained in niobium thin film cavities at 400 MHz \cite{pereira2023} and 1.3 GHz \cite{venturini2023, vegacid2023} both at around 2 K and 4 K. To further improve the promising results obtained so far, we have investigated the effect of thickness on the superconducting properties of niobium films deposited on copper using HiPIMS.

The paper is organized as follows. Section \ref{sec:level2} describes how the niobium films analyzed in this study are prepared and deposited on copper using the HiPIMS technique. In section \ref{sec:level3}, we present how we measure the superconducting properties of niobium films and investigate how these properties change with varying film thickness. In particular, the thickness effect on the critical temperature and the width of the transition from the superconducting state to the normal state is examined in section \ref{sec:level3e1}, whereas the influence of the film thickness on the lower and upper critical fields, as well as the critical current density, is explored in sections \ref{sec:level3e2} and \ref{sec:level3e3}, respectively. Finally, conclusions are drawn in section \ref{sec:level4}.

\section{\label{sec:level2}METHODS}

Niobium films of various thicknesses are deposited on copper substrates using HiPIMS. During the deposition process, this technique enhances the production of niobium ions, which can be accelerated toward the substrate. When combined with a DC bias applied to the substrate, the energy of the impinging ions increases and their trajectory is redirected close to the substrate normal, thereby enhancing the densification of the growing thin film \cite{avino2020}. 

The samples of this study are made of OFE copper with a residual resistance ratio of approximately 50 and a thickness of 2 mm. Prior to coating, the copper slabs undergo a chemical polish with a mixture of sulfamic acid (H$_{3}$NSO$_{3}$, 5 g/L), hydrogen peroxide (H$_{2}$O$_{2}$, 5\% vol.), n-butanol (5\% vol.), and ammonium citrate (1 g/L) heated to 72 $^{\circ}$C for 20 minutes. After polishing, each substrate is rinsed with sulfamic acid to eliminate any native oxide buildup and is cleaned with de-ionized water and ultra-pure ethanol. Then, substrates are placed in an ultra-high vacuum stainless steel chamber, which is subsequently connected to the coating setup. This assembly is done in an ISO5 cleanroom. Afterwards, the system is brought outside the cleanroom and connected to the pumping group and gas injection lines. The vacuum chamber is pumped down to about 10$^{-7}$ mbar. In addition, the pumping group and the entire coating system undergo a bakeout at 200 $^{\circ}$C for 48 hours. When the bakeout is completed, the vacuum chamber is maintained at 150 $^{\circ}$C until the coating process begins.

The HiPIMS deposition technique is detailed in previous works \cite{rosaz2022, arzeo2022, wang2023}, including the coating setup specifically utilized for the samples analyzed in this study. The HiPIMS discharge is 200 $\mu$s long with a repetition frequency of 100 Hz \cite{rosaz2022}. The coating temperature is monitored using an infrared thermal sensor and maintained constant at 150 $^{\circ}$C with a fan \cite{rosaz2022}. A comprehensive summary of the coating parameters is provided in table \ref{tab:table1}. Depending on the sample, ultra-pure krypton or argon is introduced into the system until a pressure ranging from 2.2$\times$10$^{-3}$ mbar to 3.4$\times$10$^{-3}$ mbar is reached. The pressure during the coating of each sample is kept constant, while the duration of the deposition varies from 1 hour to 4 hours to achieve film thicknesses ranging from approximately 1 $\mu$m to 4 $\mu$m, as the coating rate is around 1 $\mu$m/hour. However, this is a very rough estimate, as other coating parameters may also play a role. Similarly to the sputtering parameters, the voltage and current values of the plasma discharge and the DC bias to the substrate are specified for each sample in table \ref{tab:table1}. After the coating process, the samples are either maintained at a specific temperature for 48 hours, for post-coating annealing, or immediately cooled down to room temperature. The vacuum chamber is vented with dry air after this step.

\begin{table}
    \centering    
    \begin{tabular}{| c || c | c | c | c | c | c | c | c |}
    \hline
    \multicolumn{1}{|c||}{\pmb{Sample}} & \multicolumn{3}{c|}{\pmb{Sputtering parameters}} & \multicolumn{2}{c|}{\pmb{DC bias}} & \multicolumn{2}{c|}{\pmb{Plasma discharge}}  & \multicolumn{1}{c|}{\pmb{After coating}}\\
    \cline{2-9}
    
      \pmb{}    &  \pmb{Gas} & \pmb{Pressure} (Pa) & \pmb{Duration} (min) & \pmb{Voltage} (V) &  \pmb{Current} (A) &   \pmb{Voltage} (V) &  \pmb{Current} (A) &  \pmb{Temperature} (°C)\\ [0.5ex]
     \hline 
       R8-A2  & Ar & 0.23 & 180 & -50 & 47 & 590 & 160 & $-$ \\
         \hline
       R8-D2  & Ar & 0.23 & 60 & -50 & 47 & 590 & 160 & $-$ \\
         \hline
       R11-C2  & Ar & 0.26 & 240 & -75 & 60 & 585 & 180 & $-$ \\
         \hline
       R16-A1  & Kr & 0.26 & 180 & -75 & 50 & 625 & 162 & 200 \\
         \hline
       R16-B2  & Kr & 0.26 & 120 & -75 & 50 & 625 & 162 & 200 \\
         \hline
       R16-C1  & Kr & 0.26 & 240 & -75 & 50 & 625 & 162 & 200 \\
         \hline
       R17-A1  & Kr & 0.27 & 240 & -75 & 51 & 620 & 158 & 250 \\
         \hline
       R17-C2  & Kr & 0.27 & 120 & -75 & 51 & 620 & 158 & 250 \\
         \hline
       R18-B1  & Kr & 0.24 & 120 & -75 & 54 & 620 & 162 & 300 \\
         \hline
       R18-C2  & Kr & 0.24 & 240 & -75 & 54 & 620 & 162 & 300 \\
         \hline
       R19-A1  & Kr & 0.22 & 240 & -75 & 59 & 630 & 162 & 150 \\
         \hline
       R19-B2  & Kr & 0.22 & 60 & -75 & 59 & 630 & 162 & 150 \\
         \hline
       R21-C2  & Kr & 0.34 & 240 & -75 & 36 & 645 & 135 & 350 \\
         \hline
    \end{tabular}
    \caption{Coating parameters for the niobium films deposited on copper using the HiPIMS technique. The table contains information on the sputtering gas, the pressure in the vacuum chamber during deposition, and the duration of the coating expressed in minutes. Moreover, the table includes values for the voltage and current associated with the DC bias and the plasma discharge. Finally, the temperature at which the sample is maintained for 48 hours post-deposition is specified. If the temperature is not specified in the table, the sample is cooled to room temperature after the coating process.}
    \label{tab:table1}
\end{table}



Before measuring the superconducting properties, all samples are cut to dimensions of approximately 3$\times$3 mm$^{2}$ using electro-erosion. In addition, the thickness of each sample is measured by X-ray fluorescence via the attenuation method. Sample dimensions are essential for determining certain superconducting properties of the samples, as will be further explained in section \ref{sec:level3}.

\section{\label{sec:level3}RESULTS AND DISCUSSION}

Superconducting properties of niobium films deposited on copper substrates are measured using a Superconducting Quantum Interference Device Vibrating Sample Magnetometer (SQUID-VSM) \cite{manual_VSM}. The sample is vibrated at a known frequency and phase-sensitive detection is employed to enhance data collection speed and reject spurious signals. Unlike traditional vibrating sample magnetometers, the signal generated by a sample is not influenced by the frequency of vibration but is exclusively governed by the magnetic moment of the sample, the vibration amplitude, and the configuration of the SQUID detection circuit \cite{manual_VSM}. The setup includes a superconducting coil that can generate magnetic fields of up to 7 T, while also providing precise temperature control within the range of 1.8 K to 400 K. The external magnetic field is applied parallel to the plane of niobium films. 

For all measurements presented in this study, the temperature sweep rate for determining the superconducting properties is kept constant at 1 K per minute throughout the data collection, whereas the field sweep rate for measuring the magnetization curve is kept equal to 10 mT/s. The sample undergoes a zero-field cooling prior to each measurement. Furthermore, to remove any residual magnetic flux potentially trapped in the superconducting coil, the coil is quenched before initiating the zero-field cooling of the sample. 

\begin{figure}[!htb]
   \centering
   \includegraphics*[width=0.55\columnwidth]{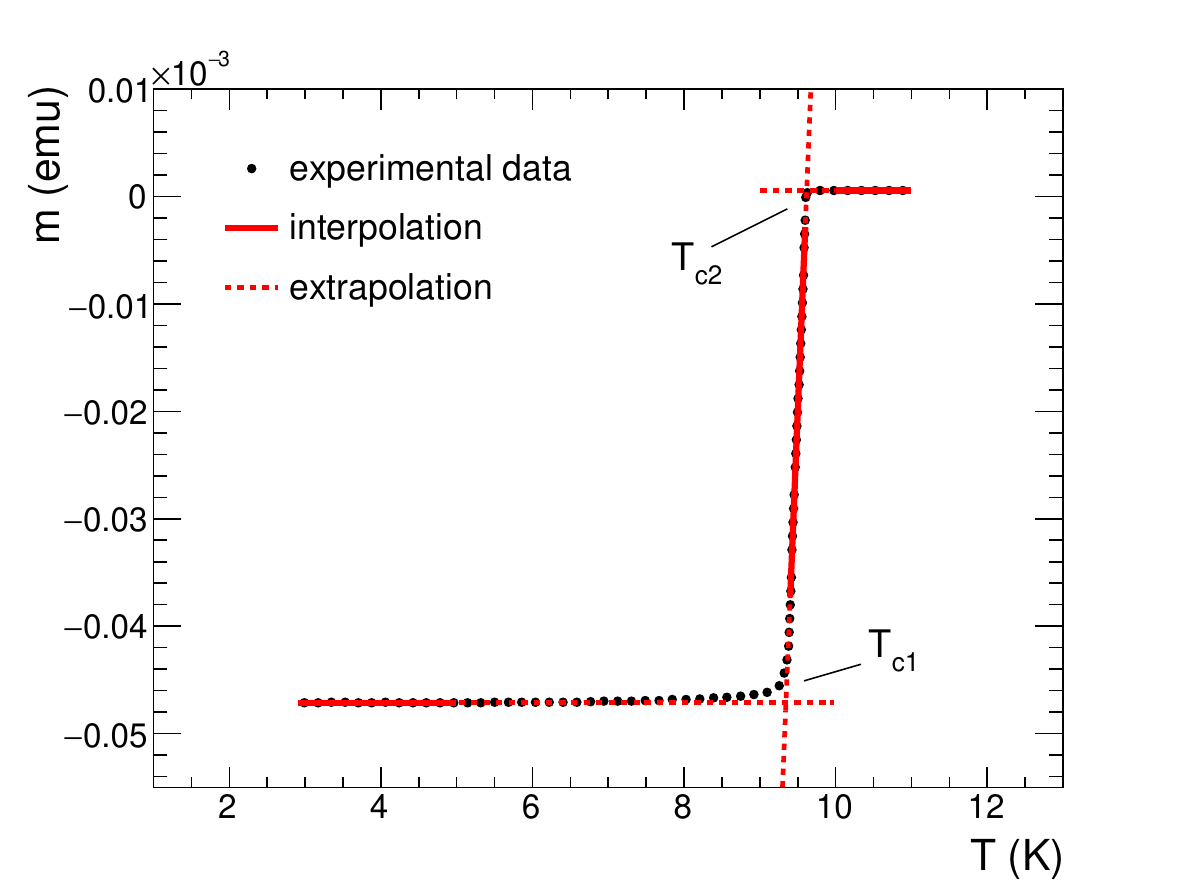} 
   \caption{Magnetic moment ($m$) value of sample R11-C2 as a function of temperature ($T$) in the presence of an applied magnetic field of 1.2 mT. The interpolation of data in different temperature ranges is represented by solid lines, while the extrapolation of the fitted curves is displayed by dashed lines.}
   \label{fig:plot1}
\end{figure}

Superconducting properties of samples are examined by measuring the magnetic moment of the samples at various values of temperature and applied magnetic field. More precisely, we measured the magnetic moment $m$ of each sample for different values of the applied magnetic field from approximately 2.5 K up to about 12 K. The transition of the sample from the superconducting state to the normal state corresponds to a rather abrupt increase in the magnetic moment. The critical temperature and the transition width of each sample are determined by measuring the magnetic moment in the magnetometer as a function of temperature, with the application of an external magnetic field of 1.2 mT. This low magnetic field value is chosen to enable the measurement of these superconducting properties in the presence of an almost negligible magnetic field. As shown in figure \ref{fig:plot1}, the onset of the change in the magnetic moment resulting from the transition of the sample from the superconducting state to the normal state is interpolated with a straight-line segment. Similarly, the magnetic moment values from 2.5 K to 5 K and from 9.6 K to 11.5 K are interpolated with horizontal lines. By identifying the intersection point of the fitted curves as shown in figure \ref{fig:plot1}, it is possible to determine the temperatures $T_{c1}$ and $T_{c2}$ that define the transition width of the sample from the superconducting state to the normal state. The value of critical temperature $T_{c}$ is calculated by averaging the temperatures $T_{c1}$ and $T_{c2}$, while the transition width $\Delta T_{c}$ is equal to the difference between $T_{c1}$ and $T_{c2}$.

The magnetization curve of each sample is assessed at various temperatures approximately between $-$2 T and 2 T. The magnetic moment of each sample is measured as the external magnetic field is gradually increased from 0 T to 2 T, then reversed from 2 T to $-$2 T, and finally increased again from $-$2 T to 2 T. At the beginning of each magnetization curve measurement, the initial measurements of the magnetic moment as the applied magnetic field is progressively increased from the value of 0 T determine what is known as the virgin magnetization curve. The virgin magnetization curve initially exhibits a linear trend due to the perfect repulsion of the external magnetic field by the superconducting sample. Interpolating the virgin magnetization curve with a linear function enables the measurement of the entry field. Indeed, the entry field $B_{entry}$ is identified when the magnetic moment starts to deviate from the initial linear behavior observed in the virgin magnetization curve. In this work, the entry field is defined as the applied magnetic field value where the discrepancy between the measured magnetic moment of the sample and the linear fit of the low-field data exceeds one standard deviation. The entry field is correlated with the lower critical field $B_{c1}$. 
In section \ref{sec:level3e2}, we make the assumption that $B_{c1}$ equals $B_{entry}$ solely for the purpose of evaluating the Ginzburg-Landau parameter of each sample. This assumption is valid in the absence of pinning and if no sample shape factors are involved. Some precautions have been taken to minimize potential limitations arising from our assumption. Each sample is cooled down in a zero-applied magnetic field before each measurement to minimize the pinning of the external magnetic field. In addition, we attempted to measure samples with approximately the same dimensions to minimize significant variations in the data due to their shape. Despite these precautions, a small discrepancy may still exist between $B_{entry}$ and $B_{c1}$. Nevertheless, the determination of the Ginzburg-Landau parameter for our samples, where the value of $B_{c1}$ plays a role as explained in section \ref{sec:level3e2}, spans the entire range of expected values reported in the literature. This confirms that the estimation of $B_{c1}$ through the measurement of $B_{entry}$ is not far from the actual value. More details are provided in section \ref{sec:level3e2}.

\begin{figure}[!htb]
   \centering
   \includegraphics*[width=0.55\columnwidth]{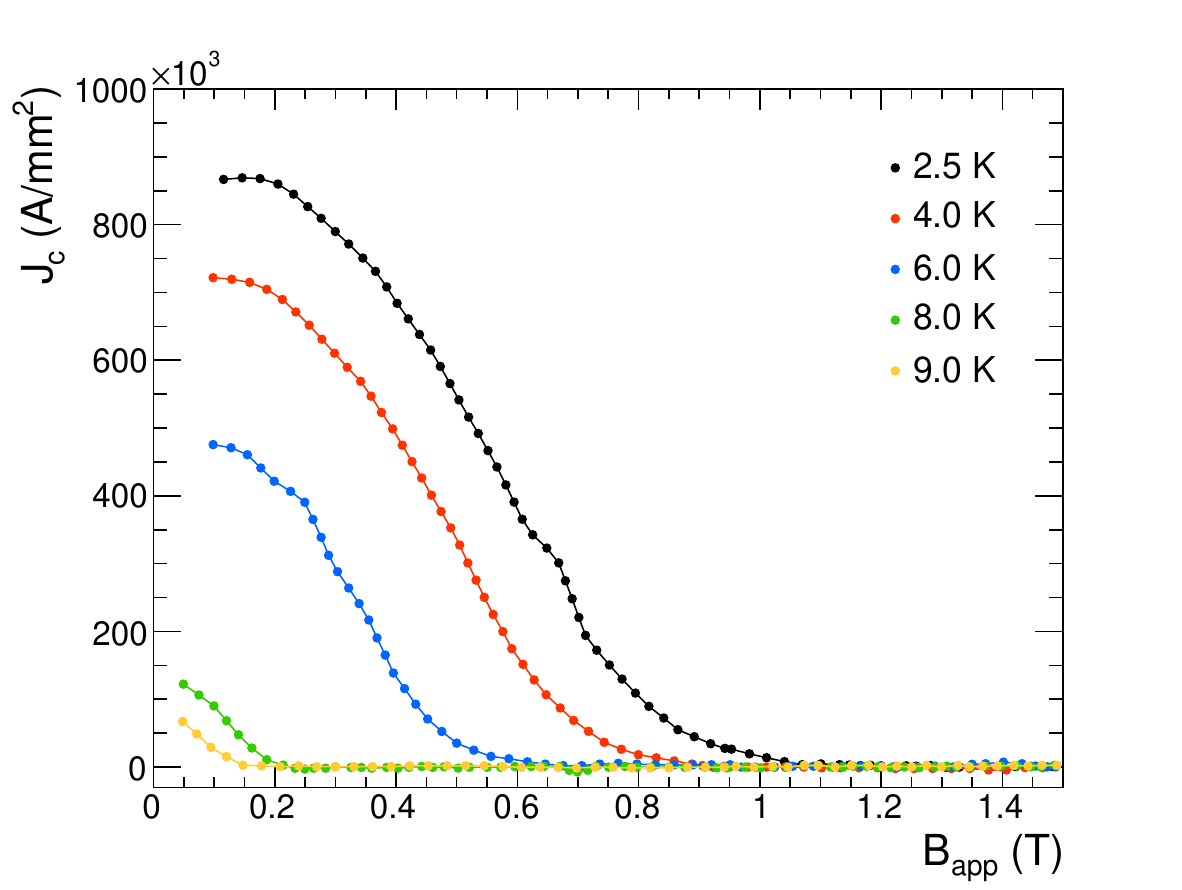} 
   \caption{Critical current density ($J_{c}$) values of sample R11-C2 as a function of the applied magnetic field ($B_{app}$) at various temperatures.}
   \label{fig:plot2}
\end{figure}

From the magnetization curves at various temperatures, we determined the temperature and magnetic field dependence of the critical current density $J_{c}$, using the Bean critical state model and applying the formula suitable for a slab geometry in the parallel magnetic field configuration \cite{bean1964, iwasa2009}:
\begin{equation}
    J_{c}(T, B_{app}) = \frac{2\Delta m(T,B_{app})}{V d}
\end{equation}
where $T$ is the temperature, $B_{app}$ is the applied magnetic field, $\Delta m(T,B_{app})$ represents the separation between the two branches of the magnetic moment loop measured with opposite field sweep directions, $V$ is the volume of the sample and $d$ its thickness. Figure \ref{fig:plot2} shows the critical current density as a function of the applied magnetic field measured for a sample at various temperatures. For a fixed value of $B_{app}$, the lower the temperature, the higher the critical current density. We define $B_{c2}$ as the field at which $J_{c}(T, B_{app})$ becomes zero.

Hereafter, we focus on the values of $B_{c1}$ and $B_{c2}$ estimated at 4 K, because niobium-coated copper cavities are primarily used at temperatures between 4.2 K and 4.5 K. Similarly, $J_{c}$ values are assessed at 4 K and 200 mT. This specific field strength was chosen as it represents the minimum magnetic field at which the evaluation of $J_{c}$ is feasible for the majority of the samples. Table \ref{tab:table2} summarizes the superconducting properties of the niobium films on copper analyzed in this study. In particular, we report the values of $B_{entry}$, $B_{c2}$ and $J_{c}$ measured at 4 K, along with the values of $T_{c}$ and $\Delta T_{c}$ measured in the presence of an almost negligible magnetic field. Additionally, the thickness and area of each sample are provided in the same table. In this work, only samples with a $B_{entry}$ value at 4 K greater than or equal to 12 mT are considered, as those with lower values would not have significant practical applications in cavities. Indeed, radio-frequency cavities generally operate in the Meissner state, where the magnetic field is completely expelled from the superconductor that constitutes the cavities \cite{posen2015}. If we assume the value of 12 mT as the maximum peak magnetic field reached in a single-cell TESLA cavity at 1.3 GHz, it would correspond to an accelerating field of approximately 3 MV/m \cite{yamamoto2009}.

\begin{table}
    \centering
    \begin{tabular}{| c || c | c | c | c | c | c | c |}
    \hline
      \pmb{sample}    &  \pmb{d} ($\mu m$) &  \pmb{area} (mm$^{2}$) &  \pmb{$T_{c}$} (K) &  \pmb{$\Delta T_{c}$} (K) &   \pmb{$B_{entry}$} (mT) &   \pmb{$B_{c2}$} (T) &  \pmb{$J_{c}$} (A/mm$^{2}$)\\ [0.5ex]
     \hline\hline 
       R8-A2  & 3.51$\pm$0.05 & 3.05$\times$3.00 & 9.50$\pm$0.01 & 0.24$\pm$0.02 & 25$\pm$1 & 1.20$\pm$0.05  & (5.73$\pm$0.01)$\times$10$^{4}$\\
         \hline
       R8-D2  & 1.21$\pm$0.04 & 3.10$\times$2.95 & 9.39$\pm$0.02 & 0.09$\pm$0.03 & 12.5$\pm$0.5 & 1.20$\pm$0.05 & (1.26$\pm$0.01)$\times$10$^{6}$\\
         \hline
       R11-C2  & 3.54$\pm$0.09 & 2.95$\times$2.90 & 9.481$\pm$0.002 & 0.027$\pm$0.004 & 56$\pm$1 & 0.95$\pm$0.05 & (7.01$\pm$0.01)$\times$10$^{5}$\\
         \hline
       R16-A1  & 3.34$\pm$0.08 & 3.05$\times$2.95 & 9.453$\pm$0.003 & 0.103$\pm$0.006 & 19$\pm$1  & 0.40$\pm$0.05 & (6.58$\pm$0.01)$\times$10$^{4}$\\
         \hline
       R16-B2  & 2.24$\pm$0.08 & 2.98$\times$2.92 & 9.385$\pm$0.005 & 0.12$\pm$0.01 & 16$\pm$1 & 1.35$\pm$0.05 & (1.55$\pm$0.01)$\times$10$^{5}$\\
         \hline
       R16-C1  & 4.20$\pm$0.11 & 2.96$\times$2.96 & 9.463$\pm$0.003 & 0.061$\pm$0.006 & 19.5$\pm$0.5 & 0.90$\pm$0.05 & (6.95$\pm$0.01)$\times$10$^{4}$\\
         \hline
       R17-A1  & 3.82$\pm$0.13 & 3.50$\times$2.65 & 9.391$\pm$0.002 & 0.102$\pm$0.004 & 21$\pm$1 & 0.75$\pm$0.05 & (5.48$\pm$0.01)$\times$10$^{4}$\\
         \hline
       R17-C2  & 2.28$\pm$0.08 & 3.20$\times$3.10 & 9.38$\pm$0.02 & 0.22$\pm$0.04 & 21$\pm$1 & 1.55$\pm$0.05 & (9.73$\pm$0.01)$\times$10$^{4}$\\
         \hline
       R18-B1  & 2.39$\pm$0.07 & 3.15$\times$3.00 & 9.38$\pm$0.03 & 0.11$\pm$0.05 & 20$\pm$1 & 1.30$\pm$0.05 &  (9.29$\pm$0.01)$\times$10$^{4}$\\
         \hline
       R18-C2  & 3.57$\pm$0.11 & 3.00$\times$2.65 & 9.40$\pm$0.01 & 0.11$\pm$0.02 & 12$\pm$1 & 0.65$\pm$0.05 & (5.74$\pm$0.01)$\times$10$^{4}$\\
         \hline
       R19-A1  & 3.89$\pm$0.11 & 3.15$\times$2.95 & 9.423$\pm$0.006 & 0.09$\pm$0.01 & 16$\pm$1 & 0.70$\pm$0.05 & (5.65$\pm$0.01)$\times$10$^{4}$\\
         \hline
       R19-B2  & 0.96$\pm$0.03 & 3.00$\times$2.95 & 9.363$\pm$0.004 & 0.198$\pm$0.007 & 12$\pm$1 & 1.10$\pm$0.05 & (5.50$\pm$0.01)$\times$10$^{6}$\\
         \hline
       R21-C2  & 4.15$\pm$0.14 & 3.20$\times$3.15 & 9.337$\pm$0.003 & 0.128$\pm$0.005 & 13$\pm$1 & 0.70$\pm$0.05  & (4.88$\pm$0.01)$\times$10$^{4}$\\
         \hline
    \end{tabular}
    \caption{Superconducting properties of niobium films on copper analyzed in this study: $T_{c}$ is the critical temperature, $\Delta T_{c}$ is the width of the transition from the superconducting state to the normal state, $B_{entry}$ is the entry field at 4 K, and $B_{c2}$ is the upper critical field at 4 K. $J_{c}$ is the critical current density measured at 4 K and 200 mT. Moreover, the table includes the dimensions of each niobium film. The thickness of the niobium film is denoted as $d$. The width and length of each sample have a statistical error of 0.02 mm each.}
    \label{tab:table2}
\end{table}

\subsection{\label{sec:level3e1}Critical temperature and transition width}
\begin{figure}[!htb]
   \centering
   \includegraphics*[width=0.95\columnwidth]{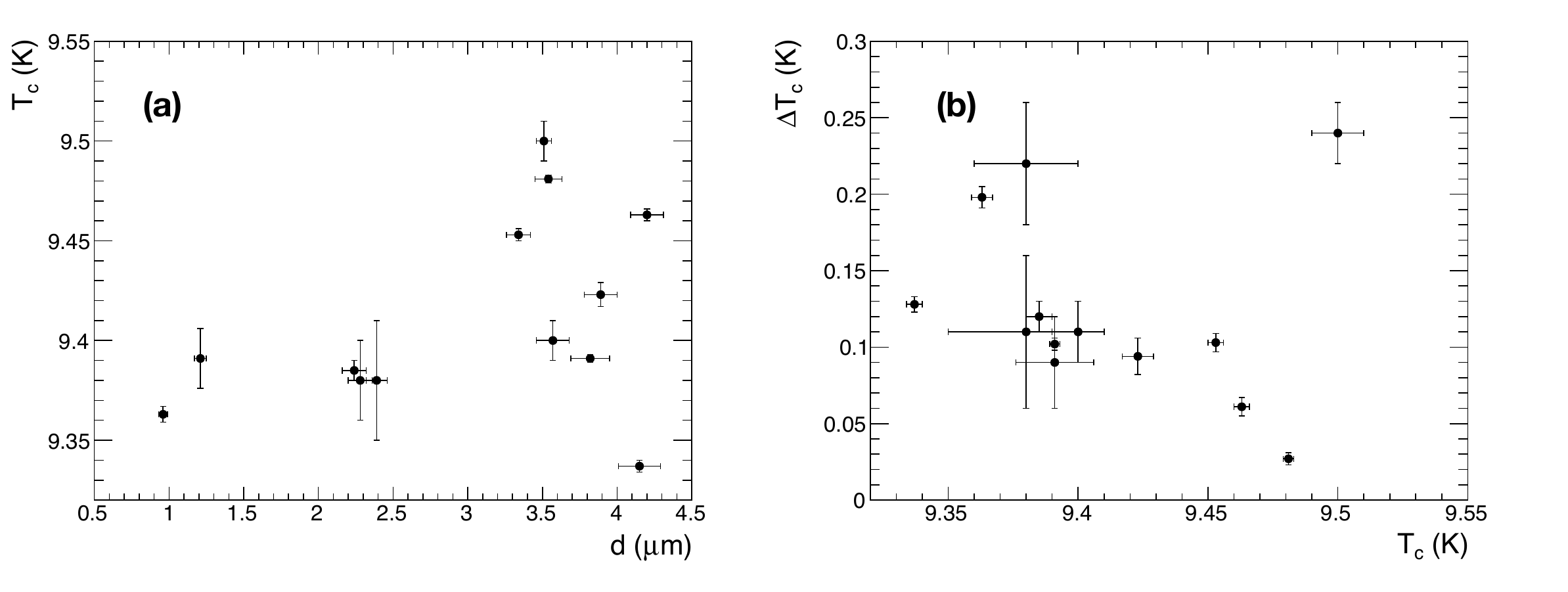} 
   \caption{(a) Dependence of the critical temperature ($T_{c}$) values on the film thickness ($d$) for niobium films on copper analyzed in this study. (b) Dependence of the transition width ($\Delta T_{c}$) on the critical temperature.}
   \label{fig:plot3}
\end{figure}

The critical temperature of niobium films has been examined in relation to film thickness. In this work, the critical temperature of all samples, ranging from approximately 9.34 K to 9.5 K, is higher than that in bulk niobium \cite{benvenuti1999, venturini2023}. Figure \ref{fig:plot3}a shows the values of critical temperature $T_{c}$ as a function of film thickness $d$ for all samples analyzed in this study. For most of the samples, the critical temperature falls between 9.34 K and 9.40 K, especially for those with a thickness lower than approximately 3 $\mu$m. On the contrary, some samples thicker than 3 $\mu$m exhibit a critical temperature exceeding 9.40 K, reaching up to 100 mK higher. In further detail, the average critical temperature of all samples with a thickness below 3 $\mu$m is about 9.38 K, whereas those thicker than 3 $\mu$m present an average critical temperature higher by 50 mK.

Figure \ref{fig:plot3}b shows the transition width $\Delta T_{c}$ as a function of the critical temperature $T_{c}$ for all samples analyzed in this study. Except for one sample with a critical temperature of 9.50 K, which significantly deviates from the observed trend, there is a general decrease in transition width as the critical temperature increases. The minimum and maximum values of the transition width differ by approximately a factor of 8. $\Delta T_{c}$ is typically correlated with the quality of the niobium film. Higher sample quality results in a narrower transition width; conversely, films with lower quality generally present a broader transition \cite{pinto2018, fonnesu2022reverse, russo2007}. 


\subsection{\label{sec:level3e2}Critical magnetic fields}

\begin{figure}[!htb]
   \centering
   \includegraphics*[width=0.95\columnwidth]{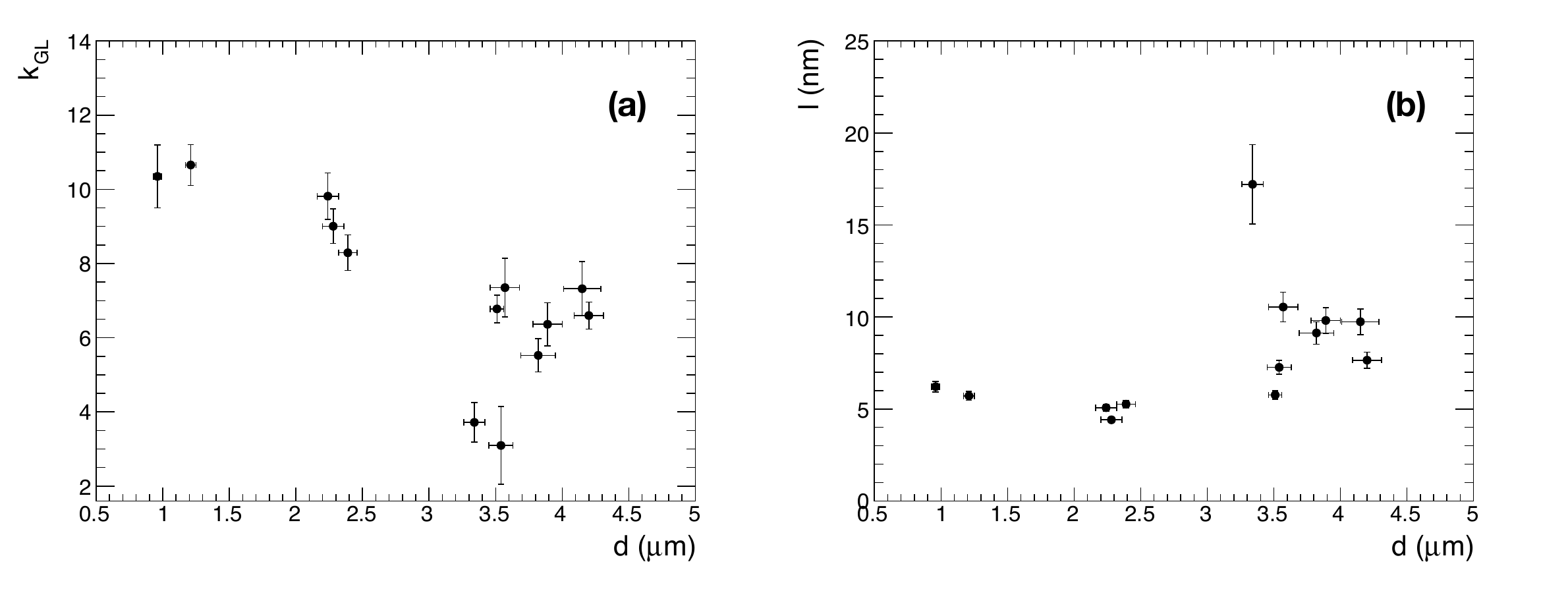} 
   \caption{(a) Dependence of the Ginzburg-Landau parameter ($k_{GL}$) on the film thickness ($d$). (b) Thickness dependence of the mean free path ($l$).}
   \label{fig:plot4}
\end{figure}

In radio-frequency cavity applications, the penetration depth and coherence length of the superconducting material, from which the cavities are made, are important parameters that determine the cavity performance. The Ginzburg-Landau parameter $k_{GL}$ is the ratio between the penetration depth $\lambda$ and the coherence length $\xi$. This parameter is also correlated to $B_{c1}$ and $B_{c2}$. Indeed, the greater the ratio between $\lambda$ and $\xi$, the larger the ratio between $B_{c2}$ and $B_{c1}$ \cite{kittel2005}. With the estimation of $B_{c1}$ and $B_{c2}$, it becomes feasible to evaluate the Ginzburg-Landau parameter of each sample. In the limit of large $\lambda$/$\xi$ values, the lower critical magnetic field $B_{c1}$ can be approximated as follows \cite{tinkham2004}:
\begin{equation}
B_{c1} = \frac{\Phi_{0}}{4 \pi \lambda^{2}}ln(k_{GL})
\end{equation}
Conversely, the upper critical magnetic field $B_{c2}$ is given by \cite{tinkham2004}: 
\begin{equation}\label{eq_Bc2}
B_{c2} = \frac{\Phi_{0}}{2 \pi \xi^{2}}
\end{equation}
Therefore, the ratio between $B_{c2}$ and $B_{c1}$ can be expressed as:
\begin{equation}\label{eq_GLparameter}
\frac{B_{c2}}{B_{c1}} = 2 k_{GL}^{2} \frac{1}{ln(k_{GL})}
\end{equation}
By obtaining the real solution of equation \ref{eq_GLparameter}, the parameter $k_{GL}$ can be determined. As explained in section \ref{sec:level3}, $B_{c1}$ has been assumed to be equal to $B_{entry}$ for estimating $k_{GL}$, although small discrepancies may be present despite precautions taken to minimize them. Another limitation arises because equation \ref{eq_GLparameter} is based on the assumption of large $\lambda$/$\xi$ values, which could potentially lead to uncertainties in the estimation of $k_{GL}$, especially for niobium films with $k_{GL}$ values close to those of bulk niobium. Nevertheless, the $k_{GL}$ values of our samples span almost the entire range expected for niobium films used in thin film radio-frequency cavity applications \cite{padamsee2}. Figure \ref{fig:plot4}a shows the Ginzburg-Landau parameter of all samples at 4 K as a function of the film thickness. Since the values of $B_{entry}$ and $B_{c2}$ listed in table \ref{tab:table2} refer to a temperature of 4 K, $k_{GL}$ is the Ginzburg-Landau parameter evaluated at 4 K. Our niobium films demonstrate $k_{GL}$ values within the range of 3 to 11, while $k_{GL}$ values of niobium films for superconducting cavities typically vary from 3.5 to 12 \cite{padamsee2}. Obtaining $k_{GL}$ values within the expected range enables us to have confidence in the Ginzburg-Landau parameter determination procedure employed in this work. Minor deviations from the actual value, resulting from assuming $B_{c1}$ equal to $B_{entry}$ and the partial fulfillment of the large $\lambda$/$\xi$ limit in some samples, have negligible implications for the scope of this study. As shown in figure \ref{fig:plot4}a, a correlation between the Ginzburg-Landau parameter and the film thickness has been observed. In particular, the Ginzburg-Landau parameter generally tends to decrease with increasing film thickness. This may suggest that increasing the thickness of the niobium film on a copper cavity leads to lower values of surface resistance. It is worth mentioning that two samples, with a thickness of approximately 3.4 $\mu$m, exhibit a $k_{GL}$ value of around 3, which is quite close to the lower limit of the expected range in niobium films for radio-frequency cavities. This does not appear to be correlated with different coating parameters when compared to all other samples in the study. Furthermore, one sample was produced using argon, and the other using krypton.

\begin{figure}[!htb]
   \centering
   \includegraphics*[width=0.55\columnwidth]{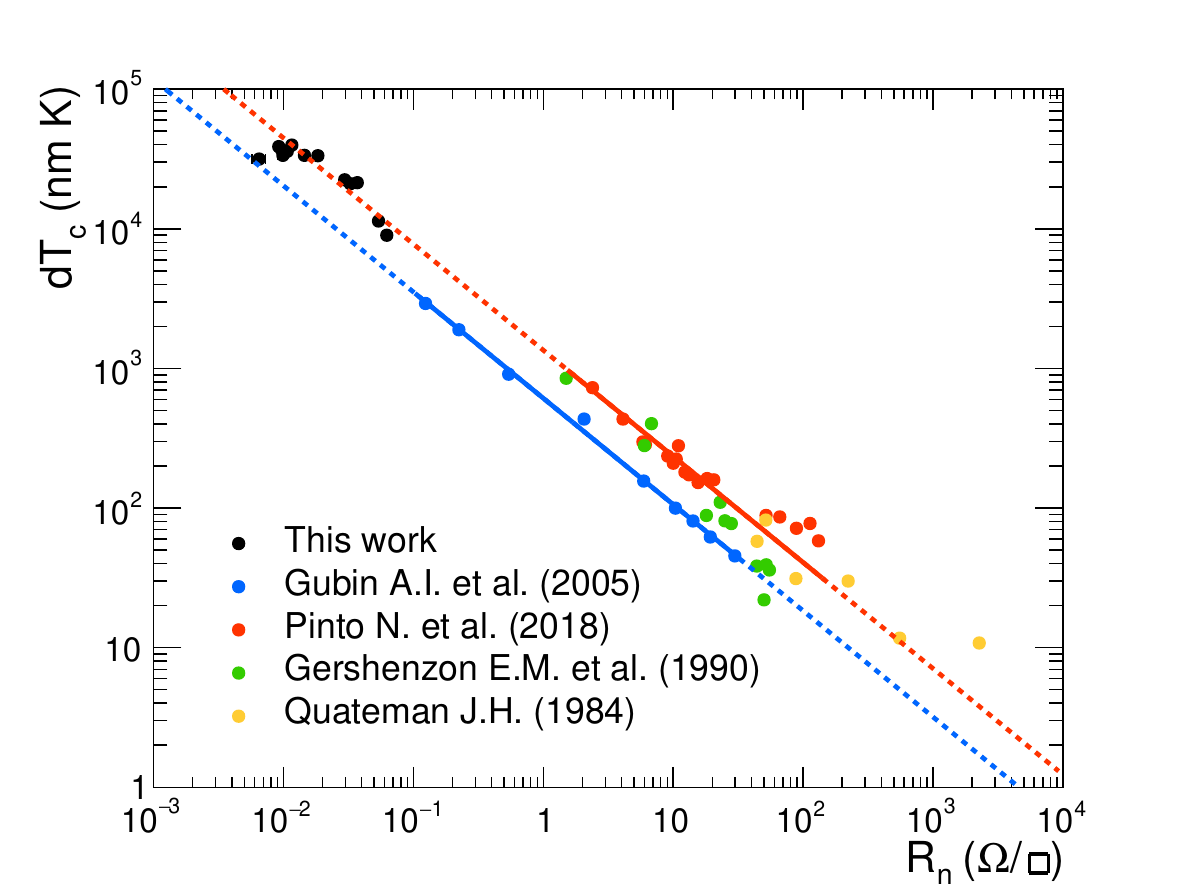} 
   \caption{Dependence of the product of the film thickness ($d$) and the critical temperature ($T_{c}$) on the sheet resistance at the normal state ($R_{n}$) for niobium films. Data from various datasets are reported: in blue, the data by Gubin et al. \cite{gubin2005}, in red, the data by Pinto et al. \cite{pinto2018}, in green and yellow, those by Gershenzon et al. \cite{gershenzon1990} and Quateman et al. \cite{quateman1986}, respectively. Data in black corresponds to this study. In addition, the two trends derived by Ivry et al. \cite{ivry2014} from the data of Gubin et al. and by Pinto et al. from their own data are shown. The solid lines correspond to the interpolation range, while the dashed lines indicate the extrapolation of the observed trends.}
   \label{fig:plot5}
\end{figure}

In the dirty limit, the coherence length as a function of temperature can be expressed as \cite{orlando1979}:
\begin{equation}
\xi(T) = 0.852 \sqrt{\xi_{0} l} \frac{1}{\sqrt{1-\frac{T}{T_{c}}}}
\end{equation}
where $l$ is the mean free path and $\xi_{0}$ is the BCS coherence length, which is 38 nm in niobium \cite{kittel2005}. By substituting the expression of the coherence length into equation \ref{eq_Bc2}, one obtains the mean free path $l$ of each sample:
\begin{equation}
l = \frac{\Phi_{0}}{2 \pi B_{c2}} \frac{1}{(0.852)^{2} \xi_{0}}\left(1-\frac{T}{T_{c}}\right)
\end{equation}
where $T$ is assumed to be equal to 4 K as $B_{c2}$ is evaluated at that temperature. Figure \ref{fig:plot4}b shows the values of the mean free path as a function of the film thickness. The values of $l$, which span from $\sim$4 nm to $\sim$17 nm for the samples analyzed in this study, are very close to the expected range where the surface resistance of niobium thin film cavities exhibits a minimum \cite{bonin1996, benvenuti1999}. The dirty limit assumption may not be entirely satisfied, as the values of $l$ in our samples are smaller than $\xi_{0}$ only by a factor ranging from 2 to 9, whereas the dirty limit condition requires that the value of $l$ be significantly smaller than $\xi_{0}$ \cite{orlando1979}. Ivry et al. discovered a universal relation among the critical temperature, the film thickness, and the sheet resistance of the film at the normal state ($R_{n}$) \cite{ivry2014}. This relation has been demonstrated to be valid for various superconducting materials across a wide range of thicknesses (approximately from 1 nm to 1 $\mu$m) \cite{ivry2014}:
\begin{equation}
d T_{c} = A R_{n}^{-B}
\end{equation}
where $A$ and $B$ are fitting parameters. Ivry's relation is valid for superconducting films deposited on any substrates. In accordance with the convention adopted in the original paper \cite{ivry2014}, $d$, $T_{c}$ and $R_{n}$ are considered unitless when the values provided are in nanometers, Kelvin, and ohms per square, respectively. Similarly, $A$ and $B$ are hereafter considered dimensionless. We have used Ivry's relation to validate our mean free path estimation for the samples analysed in this work. Indeed, although the sheet resistance of our films in the normal state has not been directly measured, it can be estimated using the relation between electrical resistivity $\rho_{0}$ and mean free path:
\begin{equation}
\rho_{0} l = \textrm{constant}
\end{equation}
where the material constant of niobium is 3.72$\times$10$^{-6}$ $\mu \Omega$ cm$^{2}$ \cite{mayadas1972}. As a consequence, $R_{n}$ is equal to $\rho_{0}/d$. To verify the scaling relation in niobium, Ivry et al. utilized data from Gubin et al. \cite{gubin2005}, which covered niobium films with thicknesses ranging from 7 nm to 300 nm. They found that $A$ is equal to 611.38 and $B$ is equal to 0.761 in niobium films \cite{ivry2014}. More recently, Pinto et al. \cite{pinto2018} found in a different dataset that the parameters $A$ and $B$ are 1350$\pm$120 and 0.76$\pm$0.05, respectively. The fitting parameter $B$ is in excellent agreement in both studies, while the parameter $A$ in Pinto's study is larger than that in Ivry's. Figure \ref{fig:plot5} shows the two different trends derived from their respective datasets. Additional data from the datasets of Gershenzon et al. \cite{gershenzon1990} and Quateman et al. \cite{quateman1986} appear to mainly follow the trend recently found by Pinto et al. Similarly, our data primarily follow the same trend observed in the datasets from Pinto, Gershenzon, and Quateman, as shown in figure \ref{fig:plot5}. This suggests that the derivation of the mean free path for niobium films analysed in this work is reasonably satisfactory, even if the dirty limit assumption is not fully met in some samples. In addition, the relation among $d$, $T_{c}$, and $R_{n}$ found by Ivry et al. appears to be valid not only for niobium films with a thickness lower than 300 nm but also within the range of $\sim$1 to $\sim$4 $\mu$m, which corresponds to the range of our samples.

\subsection{\label{sec:level3e3}Critical current density}

\begin{figure}[!htb]
   \centering
   \includegraphics*[width=0.55\columnwidth]{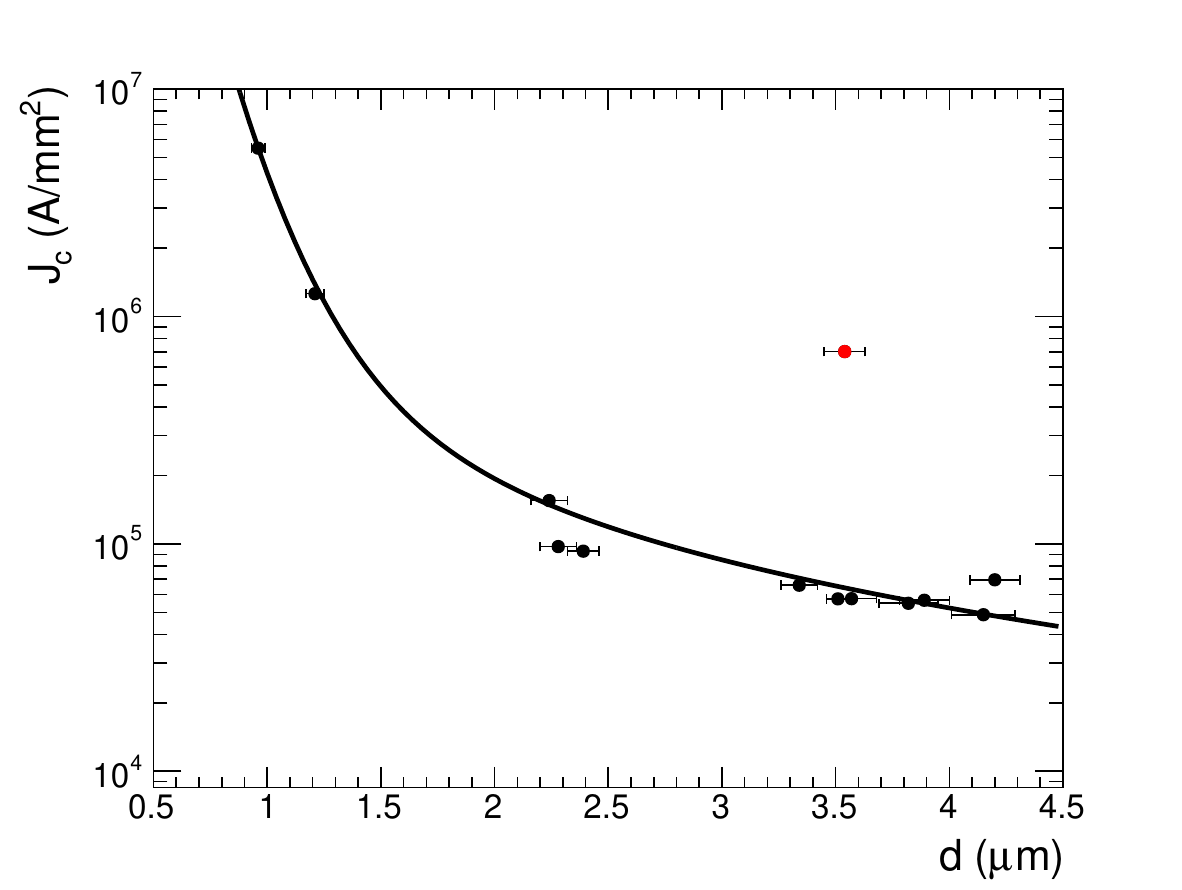} 
   \caption{Values of critical current density ($J_{c}$) at 4 K and 200 mT as a function of the film thickness ($d$). Except for the sample in red which significantly deviates from the observed trend, the data are interpolated from the sum of two power functions.}
   \label{fig:plot6}
\end{figure}

The critical current density $J_{c}$ can be significantly enhanced through the creation of pinning centers that effectively anchor fluxons against the Lorentz force acting on them. These pinning centers are created within the crystalline lattice and can result from material impurities, dislocations, grain boundaries, and precipitates with sizes of the order of the coherence length \cite{iwasa2009, dhavale2012flux}. For example, it is well-known that metallurgical processes, such as cold working or heat treatment, typically increase the density of pinning centers and, as a result, enhance the critical current density in type-II superconductors \cite{iwasa2009}. This is also applicable to niobium, where the critical current density can provide insights into the presence of structural defects within the crystalline lattice \cite{tedmon1965, antoine2019, venturini2023}. In this study, the critical current density of niobium films has been determined using the Bean critical state model in a slab geometry under the influence of a parallel magnetic field. The values of $J_{c}$ at 4 K and 200 mT for all films are provided in table \ref{tab:table2}, while the thickness dependence of these values is presented in figure \ref{fig:plot6}. 

Except for one data point that significantly deviates from the observed trend, the data in figure \ref{fig:plot6} are interpolated from the following function:
\begin{equation}
J_{c} = (3.8\pm0.7) \times d^{-(6.8\pm0.8)} + (0.50\pm0.08) \times d^{-(1.6\pm0.1)}
\end{equation}
where $J_{c}$ and $d$ are given in MA/mm$^{2}$ and $\mu$m, respectively. The decrease in $J_{c}$ values with increasing film thickness indicates a progressive reduction in the density of defects and structural irregularities in the crystalline lattice as the film thickness increases. A niobium film with a thickness of 4 $\mu$m exhibits a critical current density of approximately 5$\times$10$^{4}$ A/mm$^{2}$. However, this value is still higher than that of bulk niobium. Based on the interpolation of $J_{c}$, a thickness of 11 $\mu$m results in a $J_{c}$ value of 10$^{4}$ A/mm$^{2}$, which is two orders of magnitude higher compared to that of niobium ingots used in the production of bulk niobium radio-frequency cavities \cite{dhavale2012flux}.

\section{\label{sec:level4}CONCLUSION}

The purpose of this study is to understand how the superconducting properties of niobium films deposited on copper using HiPIMS vary with changing film thickness. The HiPIMS coating technique has demonstrated promising results in mitigating the $Q$-slope and reducing the residual surface resistance in niobium-coated copper radio-frequency cavities. Therefore, it holds the potential for successful application in the production of cavities for upcoming particle accelerators.

In this study, we analyzed the superconducting properties of niobium films with thicknesses ranging from approximately 1 $\mu$m to 4 $\mu$m. The results presented here indicate that both the critical temperature and the Ginzburg-Landau parameter, calculated from the ratio between the lower and upper critical magnetic fields, show a correlation with the film thickness. In general, a high critical temperature and a small Ginzburg-Landau parameter contribute to reducing the surface resistance of superconductors in the radio-frequency regime. In the pursuit of optimizing the surface resistance of niobium films for radio-frequency cavity applications, increasing the film thickness appears to align with this objective. Indeed, thicker niobium films, analysed in this study, exhibit a higher critical temperature and a smaller Ginzburg-Landau parameter. Moreover, both transition width and critical current density decrease with increasing thickness; therefore, the quality of the films appears to improve, while the presence of defects and structural irregularities within the crystalline lattice seems to decrease, which is desirable for radio-frequency cavity applications. Finally, in this study, we assessed the mean free path of our samples in the dirty limit based on the experimental measurements of the upper critical magnetic field. The obtained values of the mean free path are in quite close proximity to the expected value where the surface resistance of niobium thin film radio-frequency cavities exhibits a minimum. Using the mean free path estimation in our samples, the scaling relation among critical temperature, thickness, and sheet resistance at the normal state, as discovered in superconducting films by Ivry et al., appears applicable to niobium films up to approximately 4 $\mu$m. This extends the previously confirmed validity for niobium films, which was limited to around 300 nm thickness.

Based on our findings, thicker niobium films deposited on copper using the HiPIMS technique demonstrate improved superconducting properties and reduced densities of defects and structural irregularities in the crystalline lattice. Consequently, these improvements hold promise for enhancing overall performance and potentially mitigating the $Q$-slope observed in niobium thin film radio-frequency cavities. Nevertheless, finding a compromise between film thickness and structural aspects in cavities is essential. This involves addressing concerns like the potential detachment of the film from the substrate, which is more likely with thicker films, mechanical issues caused by differing thermal contractions between niobium films and copper substrates, and ensuring uniform coating of films with an adequate thickness across extensive areas.


\section*{\label{sec:level6}ACKNOWLEDGEMENTS}
The authors would like to thank Dr. Dorothea Fonnesu and Dr. Akira Miyazaki for useful discussion. The research leading to this work is part of the Future Circular Collider study.






\renewcommand{\bibsection}{\subsection*{REFERENCES}}

\bibliography{apssamp}

\end{document}